\documentclass[epj,nopacs,fleqn]{svjour}
\pdfoutput=1
\usepackage{graphicx,amsmath,amssymb,slashed,cite,verbatim,url}
\usepackage{color}
\usepackage{courier}
\newcommand{\code}[1]{\texttt{#1}}

\newcommand{\nn}{\nonumber}

\newcommand{\tn}{\tabularnewline}
\newcommand{\ct}{c_{\theta}}
\newcommand{\st}{s_{\theta}}

\newcommand{\rosetta}{\textsc{Ro\-set\-ta}}
\newcommand{\feynrules}{\textsc{FeynRules}}
\newcommand{\massbasis}{BSMC}
\newcommand{\be}{\begin{equation}}
\newcommand{\ee}{\end{equation}}
\def\bsp#1\esp{\begin{split}#1\end{split}}

\begin{document}

\title{\rosetta: an operator basis translator for Standard Model effective field theory}

\author{
 Adam Falkowski\inst{1},
 Benjamin Fuks\inst{2},
 Kentarou Mawatari\inst{3},
 Ken Mimasu\inst{4},
 Francesco Riva\inst{5},
 Ver\'onica Sanz\inst{4}
}

\institute{
  Laboratoire de Physique Th\'eorique, Bat.~210,
  Universit\'e Paris-Sud, 91405 Orsay, France
  \and
 Institut Pluridisciplinaire Hubert Curien/D\'epartement Recherches
 Subatomiques, Universit\'e de Strasbourg/CNRS-IN2P3,
 23 rue du Loess, F-67037 Strasbourg, France
 \and
 Theoretische Natuurkunde and IIHE/ELEM, Vrije Universiteit Brussel,
 and International Solvay Institutes,\\
 Pleinlaan 2, B-1050 Brussels, Belgium
 \and
 Department of Physics and Astronomy, University of Sussex, 
 Brighton BN1 9QH, UK
 \and
 CERN, Theory Division, 1211 Geneva, Switzerland.
}

\abstract{
  We introduce \rosetta, a program allowing for the translation between different bases of effective
  field theory operators. We
  present the main functions of the program and provide an example of usage.
  One of the Lagrangians which \rosetta\ can translate into has been implemented into
  \feynrules, which allows \rosetta\ to be interfaced into various
  high-energy physics programs such as Monte
  Carlo event generators. In addition to popular bases choices, such as the Warsaw and
  Strongly Interacting Light Higgs bases already implemented in the program,
  we also detail how to add new operator bases into the \rosetta\ package. In this way, phenomenological studies using an effective
  field theory framework can be straightforwardly performed.}

\date{}

\newcommand{\adam}[1]{{\color{magenta} #1}} 
\newcommand{\km}[1]{{\color{blue} #1}} 

\titlerunning{Rosetta}
\authorrunning{Falkowski, Fuks, Mawatari, Mimasu, Riva and Sanz}

\maketitle


\vspace*{-13cm}
\noindent 
\small{MCNET-15-25}
\vspace*{11.2cm}

\section{Introduction}\label{sec:intro}
The start of a second LHC experimental era raises new hopes to detect physics
beyond the Standard Model (BSM). The high energy of the experiment increases the
chances of a direct discovery of new physics resonances, while a combination of
high energy and high luminosity favors the possible observation of new phenomena via Standard Model (SM) precision tests. Interestingly the latter offers a
complementary and model-independent tool for BSM searches if the results are
interpreted in the context of an effective field theory (EFT). The EFT indeed
captures in a general way the low-energy effects of heavy new physics from a
bottom-up perspective. More precisely, it systematically organizes possible
departures from the SM as an expansion in the energy at which the
processes of interest occur over the (high) new physics scale, and
simultaneously provides a dictionary to interpret these departures in the
context of explicit BSM models.

Given the SM field content (including a single Higgs doublet), assuming baryon
and lepton number conservation, flavor universality and a linear realization of
the electroweak symmetry, the leading effects implied by an EFT description
consist of dimension-six operators that are supplemented to the SM Lagrangian.
At this order, 59 (76 real) new
independent coefficients~\cite{Buchmuller:1985jz,Grzadkowski:2010es}\footnote{Relaxing flavor universality, the number of
independent dimension-six operators grows to 2499~\cite{Alonso:2013hga}.} capture all
possible deformations from the SM. Despite this large number of new free
parameters, important classes of observables (\textit{e.g.}, Higgs production
and decay or $Z$-pole observables) depend on a much smaller subset of
parameters~\cite{Pomarol:2013zra,Ellis:2014dva,Gupta:2014rxa,Ellis:2014jta,%
Falkowski:2014tna,Efrati:2015eaa}.
Owing to that, the EFT approach is not only useful for parameterizing BSM
searches but is also testable \textit{per se} by looking at correlations among
the expected signatures.

Another important aspect of the EFT approach is the choice of the operator basis,
so that a given physical effect could be modeled by different combinations of
operators at a fixed order in the EFT expansion. This well-known fact is related
to the possibility of redefining the SM fields in such a way that the zeroth
order Lagrangian in the EFT expansion (\textit{i.e.}, the SM Lagrangian) is
unaltered, while combinations of the first-order operators (\textit{i.e.},
dimension-six operators) proportional to the SM equations of motion can be
eliminated up to subleading higher-dimensional effects. For this reason,
different complete and non-redundant operator bases have been proposed in the
literature, sharing the same physical predictions but having different
advantages. The most popular choices include the so-called \emph{Warsaw}
basis~\cite{Grzadkowski:2010es}, \emph{SILH} (strongly interacting light
Higgs) basis~\cite{Giudice:2007fh,Contino:2013kra} and \emph{BSM primaries}
basis~\cite{Gupta:2014rxa,Masso:2014xra,Pomarol:2014dya}. The Warsaw basis represents the first set of non-redundant operators that has been proposed and is
particularly appropriate for
comparisons with BSM theories that modify the interactions of the SM fermions.
In contrast, the SILH basis has been designed to capture the effects of
universal theories where new physics mostly couples to the SM bosons. Finally,
the BSM primaries basis is more suitable for a bottom-up approach
since it is formulated in terms of mass-eigenstates and has a more
transparent connection to measurable quantities, its operators being aligned
with physical observables.

\renewcommand{\arraystretch}{1.5}
\begin{table*}
\center
\begin{tabular}{c|ccc}
   Basis & Underlying gauge symmetry & Fields used in the Lagrangian\\
   \hline
   Warsaw, SILH & $SU(3)_C\times SU(2)_L\times U(1)_Y$ & Gauge-eigenstates\\
   BSM primaries, Higgs & $SU(3)_C\times SU(2)_L\times U(1)_Y$ &
    Mass-eigenstates\\
   Higgs/BSM characterisation & $SU(3)_C\times U(1)_{EM}$ & Mass-eigenstates\\
\end{tabular}
\caption{Main features of the different EFT basis choices discussed in this
  document.}
\label{tab:bases}
\end{table*}
\renewcommand{\arraystretch}{1.0}

Given these multiple viewpoints, it is cumbersome to express the experimental
results in a basis-independent manner that can be readily interpreted in any of
the above-mentioned frameworks. On the other hand, different bases may be
convenient for particular applications,
either because they facilitate the comparison with a given class of
BSM theories or simply because different experimental analyses look more
transparent in a specific basis. For instance, the Warsaw basis contains an
apparent blind direction with respect to the electroweak precision tests~\cite{%
Grojean:2006nn,Gupta:2014rxa}, which introduces large theoretical correlations
among all LEP constraints. As a result, the bounds on the strength of the
dimension-six interactions appear less transparent~\cite{Han:2004az}.
The SILH basis has a similar drawback yielding a correlation between LEP2 and
LHC constraints, while the downside of the BSM primaries basis lies in the
comparison with explicit BSM models that is complicated. The \rosetta\
package that we present in this paper has been designed to explicitly solve such problems by allowing for a straightforward translation between
different EFT languages.

In addition to translating, another important goal of the \rosetta\ program is
to provide a platform for communication with Monte Carlo event generators, no
matter which EFT basis is chosen. To achieve this, we have implemented in
\rosetta\ the \emph{Higgs} basis for EFT operators that has been recently
designed by the LHC Higgs Cross Section working group (LHC\-HXSWG)~\cite{HB}.
This proposal, built on the BSM primaries basis (see Ref.~\cite{Pomarol:2014dya}), combines
two ingredients. First, all possible operators
of dimension up to six are written in terms
of the SM mass-ei\-gen\-sta\-tes 
\be
  \Delta{\cal L}^{(\rm mass)} = 
   \sum_i \frac{c_i}{\Lambda^{d_i}} {\cal O}_i (G_{\mu}^a, W_\mu^\pm, Z_\mu,
  A_\mu, h, t, b, \nu_\tau, \tau,
       \ldots) \, ,
\ee
where the operators ${\cal O}_i$ have a mass dimension ranging from two to six.
The dimensionless coefficients $c_i$ are then suppressed by an appropriate power
$d_i$ of the high-energy scale $\Lambda$, with $d_i={-2,\cdots,2}$.
We refer the ensemble of operators included in the resulting
Lagrangian, which is in spirit very similar to the Higgs characterisation
Lagrangian of Ref.~\cite{Artoisenet:2013puc}, as the \emph{BSM characterisation} (BSMC) Lagrangian.  Due to the lack of manifest $SU(2)_L\times U(1)_Y$ invariance, the
BSMC Lagrangian is associated with a larger number of independent coefficients
compared to the Warsaw, SILH or BSM primaries bases. For this reason, the second
ingredient defining the Higgs basis consists of relations among the $c_i$
coefficients that restore the full $SU(3)_C\times SU(2)_L\times U(1)_Y$
symmetry. As summarized in Table~\ref{tab:bases}, the BSM primaries and Higgs
Lagrangians are both of the form of $\Delta{\cal L}^{(\rm mass)}$, but they
additionally include constraints among the different Wilson coefficients that
render the Lagrangian invariant under the electroweak symmetry. In contrast, the
Warsaw and SILH basis Lagrangians are directly written in
terms of the SM gauge-eigenstates,
\be
  \Delta{\cal L}^{(\rm gauge)} = \sum_{i'} \frac{c'_i}{\Lambda^2}
    {\cal O}_{i'}(G_{\mu}^a, W_\mu^i, B_\mu, \Phi, Q_L, u_R, d_R,
    L_L, e_R) \, ,
\ee
and are manifestly symmetric under the electroweak symmetry group.

We have implemented the mass basis Lagrangian $\Delta{\cal L}^{(\rm mass)}$ into
\feynrules~\cite{Alloul:2013bka}%
\footnote{Implementations of the Higgs
characterisation~\cite{Artoisenet:2013puc} and the SILH
basis~\cite{Alloul:2013naa}
Lagrangians are also available.} and tuned the output format of \rosetta\ so
that the translation maps an EFT Lagrangian given in a specific basis to
$\Delta{\cal L}^{(\rm mass)}$ and generates an output file that is compatible with the
\feynrules\ implementation. As a consequence, any high-energy physics tool (and
in particular any Monte Carlo event generator) that is interfaced to \feynrules\
can be employed within the context of any EFT basis of operators that is
included in \rosetta.

With the advent of automated next-to-leading order (NLO) accurate Monte Carlo
event generation software, it is important that \rosetta\ remains flexible
enough to eventually provide compatibility with this new generation of tools.
Recent progress has been made on the theory side both in implementing the
linear dimension-six description discussed above in the \feynrules\
framework~\cite{progress} and in calculating the renormalisation group
(RG) evolution of the full set of operators and their mutual
mixing~\cite{Jenkins:2013zja,Jenkins:2013wua,Alonso:2013hga,Henning:2014wua}. In the former
case, \rosetta\ can simply be extended to provide an output compatible with the
eventual NLO model implementation, analogously to the BSMC Lagrangian. The
latter case of evaluating the RG running effects, while being a slightly
separate issue, highlights a key feature of our tool, given that the calculation of
these effects has only been performed in the original \emph{Warsaw} basis of
Ref.~\cite{Grzadkowski:2010es}. The framework provided by \rosetta\ allows for
the application of these results in any desired basis. Although the initial
version of the software does not explicitly deal with these effects, its
translation functionality can already be used in their context and we plan for future versions to incorporate RG evolution of the SM EFT Wilson coefficients.

The remainder of this paper is organized as follows. In Section~\ref{%
sec:rosetta}, we describe the basic functionalities of \rosetta\ and how to make
use of the program. Section~\ref{sec:examples} is dedicated to an example of
usage of \rosetta\ in which we focus on new physics Higgs couplings to the SM
bosons. We express them in different bases and detail the output that is
provided by \rosetta. Our work is summarized in Section~\ref{sec:conclusions}.

\section{Rosetta}\label{sec:rosetta}
The aim of \rosetta\ is to provide a modular and flexible package for EFT
basis translation and communication with event generation tools.
The primary framework which \rosetta\ has been designed to translate into is the
phenomenological effective Lagrangian,
$\Delta{\cal L}^{(\rm mass)}$, which will be explicitly defined in
Section~\ref{s:basis}.
The motivation for this choice lies in the availability of an implementation within the
\feynrules\ framework~\cite{Alloul:2013bka}, to be downloaded from the \feynrules\
model repository~\cite{FR-mb:Online},
which ensures the link with event generators and high-energy physics
programs~\cite{Christensen:2009jx,Degrande:2011ua}.

The most basic functionality of \rosetta\ is to map a chosen set of input
parameters (the Wilson coefficients in a specific basis choice)
onto the \massbasis\ coefficients such that the output can be employed
within tools relying on a BSMC basis description. As long as the input format
respects the conventions sketched in Section~\ref{s:input} and that are inspired
by the Supersymmetry Les Houches Accord (SLHA)~\cite{Skands:2003cj,%
Allanach:2008qq}, the user may define his/her own map to the \massbasis\
coefficients (or to any other basis implementation) and proceed with event
generation using the accompanying \feynrules\ implementation. This highlights
one of the
key features of \rosetta, the possibility to easily define one's own input
basis and directly use it in the context of many programs via the translation
functionality of \rosetta. The strength of this approach is that it is much simpler than
developing from scratch new modules for existing tools in the context of a 
new basis. To this end, \rosetta\ not only enables the translation
of an EFT basis into the \massbasis\ Lagrangian, but also allows for translations into any
of the other bases included in the package, \textit{i.e.}, currently the Higgs, Warsaw and
SILH bases. Translations between these three bases in any direction are
possible, so that the addition of a new basis by the user only requires the
specification of translation rules to any one of the three core bases. One is
subsequently able to indirectly translate the new basis into any of the other
two bases, as well as into the \massbasis\ Lagrangian. The details of how one can implement a new basis in \rosetta\ are discussed in Section~\ref{s:exbasis}.
\subsection{Getting started with \rosetta\label{s:commandline}}
The latest release of \rosetta\ can be obtained from\\
\verb+  http://rosetta.hepforge.org+\\
The package contains a {\sc Python} executable named \code{tran\-slate}, an
information file named \code{README} and two directories, a first folder
(named \code{Cards}) collecting example input files and a second folder (named
\code{Rosetta}) including the source code of \rosetta. The executable takes as
input an SLHA-style parameter file with the coefficients of the dimension-six
operators associated with a particular basis. Information on the format of such
an input file can be found in Section~\ref{s:input}. The execution of the
\code{translate} script from a shell yields the generation of an output
parameter file where all parameters are this time the coefficients of the
dimension-six operators associated with a specified new basis, the default
choice being the \massbasis\ Lagrangian.
The tool can be used by typing in
\begin{verbatim}
  ./translate PARAMCARD.dat OPTIONS
\end{verbatim}
where \code{PARAMCARD.dat} is the name of the SLHA-style input file and
\code{OPTIONS} stands for optional arguments. The latter could consist of one or
more of the following choices that will modify the behavior of the program.

\vspace*{.1cm}\noindent
\begin{tabular}{lp{5.1cm}}
    \code{-h} or \code{--help}      &This displays a help message and exits the
                                     program.\tn
    \code{-o} or \code{--output}    &This allows for the specification of the
                                     name of the output file, that is by default
                                     \code{PARAMCARD\_new.dat}.\tn
\end{tabular}

\noindent \begin{tabular}{lp{5.1cm}}
    \code{-s} or \code{--silent}    &The program suppresses warnings and takes 
                                     the default answer to any
                                     question that may have to be asked to the
                                     user.\tn
    \code{-t} or \code{--target}    &This allows for providing the name of the
                                     basis into which the translation occurs,                                        the default being \code{bsmc} and the                                           other acceptable choices being                                                  \code{higgs}, \code{silh} or \code{warsaw}.
                                     \tn
    \code{-w} or \code{--overwrite} &This allows the program to overwrite any
                                     pre-existing output file.\tn
    \code{-e} or \code{--ehdecay}   &This allows to use the interface with the
                                     e{\sc HDecay}
                                     program~\cite{Contino:2013kra} for the
                                     calculation of the Higgs boson width and
                                     branching fractions. See
                                     Section~\ref{ss:ehdecay}.\tn
    \code{-f} or \code{--flavor}   &This allows to specify the treatment of the
                                     flavor structure relevant for the fermionic
                                     operators, the default being
                                     \code{general} and the other acceptable
                                     choices being \code{universal}
                                     and \code{diagonal}. See 
                                     Section~\ref{ss:flavour}.\tn
    \code{-d} or \code{--dependent} &This allows the program to also write out 
                                     any dependent parameters calculated by the 
                                     translation function to the output file.
\end{tabular}
\vspace*{.1cm}

On run time, \rosetta\ starts by performing several checks on the input
parameters and verifies the consistency of the input file with respect to the
specifications of the internal basis implementation. In this way, any missing SM
inputs (with respect to the requirements included in the \code{required\_inputs}
and \code{required\_masses} attributes of the basis class, see
Section~\ref{s:struct}) can be included using the value provided in the
Particle Data Group (PDG) review~\cite{Agashe:2014kda}, while any missing
coefficient associated with an operator that is present in the basis (and thus
declared in the \code{independent} attribute of the basis class, see
Section~\ref{s:struct}) can be included with a zero value.

Once the translation is achieved, \rosetta\ outputs a new parameter file that is
by default named \code{PARAMCARD\_new.dat}. This file contains the values of all
parameters relevant for the target basis and also includes the necessary
modifications to the input parameters, such as the $W$-boson mass that may
depend on some dimension-six operator coefficients.
\subsection{Input files and their handling in \rosetta\label{s:input}}
\rosetta\ requires input parameters to be given under the form of a file encoded
in a format similar to the SLHA one detailed in Refs.~\cite{Skands:2003cj,%
Allanach:2008qq}. Parameters are grouped into blocks and each
parameter is identified inside its own block by one or more integer numbers
called counters. For instance, the SM inputs necessary for the definition of the
SILH basis would read
\begin{verbatim}
  BLOCK SMINPUTS #
    1    +1.27916e+02 # aEWM1
    2    +1.16638e-05 # Gf
    3    +1.18400e-01 # aS
    4    +9.11876e+01 # MZ
    25   +1.25000e+02 # MH
\end{verbatim}
where the different entries respectively correspond to the inverse of the
electromagnetic coupling constant (\code{aEWM1}), the Fermi constant
(\code{Gf}), the strong coupling constant (\code{aS}), the $Z$-boson mass
(\code{MZ}) and the Higgs boson mass (\code{MH}). Inspired by the usual SLHA
conventions, all masses are also collected into a block called \code{MASS}
where the counters correspond to the PDG identifiers of
the particles~\cite{Agashe:2014kda}. Furthermore, matrix quantities receive a
block of their own with counters specifying the position inside the matrix.
In this way, a single block would be needed to encode, for instance, the $c_{Hud}$
coefficients associated with the ${\cal O}_{Hud}$ operator of the Warsaw basis
that is defined by
\be
  {\cal O}_{Hud} = - i \big[\bar u\gamma^\mu d\big]
      \big[\tilde\Phi^\dag D_\mu\Phi\big] \ .
\ee
In this expression, $u$ and $d$ denote the $SU(2)_L$ singlets of right-handed
up-type and down-type quark fields, respectively, and $\Phi$ and $D_\mu\Phi$
stand for the weak doublet of Higgs fields and its gauge-covariant derivative. In
flavor space, the $c_{Hud}$ coefficients take the form of a matrix, 
implemented in the input file as
\begin{verbatim}
  BLOCK WBxHud
  1 1 0.1e+00 # cHud11
  1 2 0.0e+00 # cHud12
  1 3 0.0e+00 # cHud13
  2 1 0.0e+00 # cHud21
  2 2 0.1e+00 # cHud22
  2 3 0.0e+00 # cHud23
  3 1 0.0e+00 # cHud31
  3 2 0.0e+00 # cHud32
  3 3 0.1e+00 # cHud33
\end{verbatim}
The block name contains information on the basis (\code{WB}) and on the
considered operator (\code{Hud}). Sample parameter files for all core bases can
be found in the \code{Cards} directory shipped with the program. Within those
files, we have adopted the above block naming scheme. The name of each block
starts with two letters identifying the basis (\code{BC}, \code{HB}, \code{SB}
and \code{WB} for the BSMC, Higgs, SILH and Warsaw bases respectively) that are
followed by a separator (\code{x}), and ends with the name of the
considered coefficient as it is defined in the
LHCHXSWG proposal for an EFT basis choice~\cite{HB}. In the case of EFT
operators independent of fermions, the related (non-matrix) coefficients are
collected in different blocks as a function of the Lorentz structure of the
operators. For instance, the \code{SBxV2H2} block will include all operators of
the SILH basis containing two occurrences of the Higgs field and two occurrences
of the gauge fields. Their ordering follows their order of
appearance in the LHCHXSWG proposal. The imaginary parts of all parameters are
stored in corresponding blocks whose names are prefixed with the \code{IM} tag.

The \rosetta\ package contains built-in methods for dealing with an SLHA-like
structure, and these methods have all been implemented in the
\code{Rosetta/SLHA.py} file. When an input file is read, the parser included in
the \code{SLHA.py} file recognizes the existing \code{BLOCK} and \code{DECAY}
structures of the input file and stores them as instances of the
\code{SLHA.NamedBlock}, \code{SLHA.NamedMatrix} and of the
\code{SLHA.Decay} classes. These are dictionary-like objects that can be
assigned, indexed and iterated over as regular {\sc Python} dictionaries. An
\code{SLHA.Named\-Block} object reflects the
information embedded in an SLHA block so that it possesses a \code{name}
attribute and stores values associated with integer keys as well as a mapping
from the integer keys to the parameter names. In this way, parameters can be
accessed by indexing either their integer key or their name. Similarly,
\code{SLHA.NamedMatrix} objects function analogously but operate with
a pair of indices for indexing. An
\code{SLHA.Decay} object contains an integer attribute \code{PID} that is the
PDG identifier of the particle whose decays are described by the considered
block, as well as a \code{total} attribute allowing for the storage of the total
width. Individual decay channels are then indexed by tuples of PDG codes
associated with the decay products, and the stored values are the branching
fractions. Finally, the \code{SLHA.py} file also includes the definition of the
\code{SLHA.Card} class that serves as a container for a collection of instances
of the above objects. The implementation of any basis
in \rosetta\ therefore requires the user to provide definitions for the blocks and parameters to be specified in the input file that will be read into an \code{SLHA.Card} instance belonging to that basis class. More practical information and examples are given in Sections~\ref{s:struct} and~\ref{s:exbasis}.

Three special blocks named \code{BASIS}, \code{SMINPUTS} and \code{MASS}
must always be present. The first and only element of the \code{BASIS} block
refers the name of the basis into which \rosetta\ must read the input file and
informs the program on the other blocks it should look for, based on
the structure specified in the implementation of that particular basis. This name should be a single unique string with no spaces. The
next two mandatory blocks consist of conventional input blocks specifying the
values of the SM inputs and of the particle masses. The set of required inputs will depend on the specifications in the corresponding basis implementation. Moreover, the user can
optionally specify the value of the elements of the CKM matrix by setting their
real and imaginary parts within the \code{VCKM} and \code{IMVCKM} blocks. If
absent, the information of the PDG review~\cite{Agashe:2014kda}
is used by \rosetta. All extra blocks and decay
structures are stored, left unchanged and passed to the output file unless the
user demands to use the e{\sc HDecay} program, which will overwrite any existing
decay information on the Higgs.
\subsection{Structure of \rosetta\label{s:struct}}
\renewcommand{\arraystretch}{1.2}
\begin{table}
    \begin{center}
    \begin{tabular}{cp{3.75cm}c}
        Counter  & Parameter & \rosetta\ name\tn
        \hline
        1 & The inverse of the electromagnetic coupling constant $\alpha^{-1}$
            &\code{aEWM1}\tn
        2 & The Fermi constant $G_{F}$
            &\code{Gf}\tn
        3 & The strong coupling constant $\alpha_{s}$
            &\code{aS}\tn
        4 & The $Z$-boson mass $m_Z$      &\code{MZ}\tn
        5 & The bottom quark mass $m_b$   &\code{MB}\tn
        6 & The top quark mass $m_t$      &\code{MT}\tn
        7 & The tau lepton mass $m_\tau$  &\code{MTAU}\tn
        25& The Higgs boson mass $m_H$    &\code{MH}\tn
        \hline
    \end{tabular}
    \caption{\label{t:sminputs}Identifying counters of the \code{SMINPUTS} block with the SM
       parameters allowed to be used within \rosetta. This generalizes the SLHA
       standards where the Higgs mass is ignored~\cite{Allanach:2008qq}.
       The internal names used by \rosetta\ are also given.}
    \end{center}
\end{table}
\renewcommand{\arraystretch}{1.0}

\rosetta\ is a {\sc Python} package containing the implementation of a
\code{Basis} class equipped with several utility functions for reading,
processing and writing SLHA-style parameter files. Working implementations of
bases are derived from this class and only require a small amount of information
specifying the block structure of the EFT parameters, the required SM inputs and
a series of translation functions to other existing basis implementations. In
order to be able to define a new basis class, we describe in this section the
properties of the \code{Basis} objects.

The \code{Basis} class has a number of intrinsic data members that should be
defined in order to get a working implementation of an EFT basis. These consist
of the \code{independent}, \code{required\_inputs} and \code{required\_masses}
attributes already mentioned in Section~\ref{s:commandline}, together with the
\code{name}, \code{blocks} and \code{flavored} members of the class.

\vspace*{.1cm}\noindent 
\begin{tabular}{lp{5.3cm}}
   \code{name}             &Unique string identifier for the basis
                            implementation, \emph{e.g.}, \code{higgs},
                            \code{bsmc}, \code{silh} or \code{warsaw} for
                            the core bases shipped with the package.\tn
    \code{independent}     &List of strings containing the names of the
                            independent EFT operator coefficients of the basis.
                            These are expected to be present in the input
                            parameter file.\tn
    \code{required\_inputs}&Set of integers containing the SLHA counters of the
                            required SM inputs. See Table~\ref{t:sminputs} for
                            the complete list of those allowed in \rosetta.\tn
\end{tabular}

\noindent\begin{tabular}{lp{5.3cm}}
    \code{required\_masses}&Set of integers containing the PDG identifiers of
                            the particles whose masses are required as input and
                            that are not included in \code{required\_inputs}.\tn
    \code{blocks}          &Dictionary with the non-matrix SLHA block names as keys
                            and lists of coefficients stored in that block as
                            values.\tn
    \code{flavored}       &Dictionary with matrix SLHA block names as keys.
                            The values are other dictionaries with the
                            keywords \code{kind}, \code{domain} and
                            \code{cname} as keys. This describes the properties
                            of the matrices.
\end{tabular}
In the case of the definition of a matrix block, the self-explana\-to\-ry possible
values for the keyword \code{kind} are \code{symmetric}, \code{her\-mi\-ti\-an} and
\code{general} and those related to the keyword \code{domain} are \code{real}
and \code{complex}. The properties of the ensuing matrix object will depend on the choice of these keywords. The name to be given to the individual EFT coefficients are
derived from the value of the keyword \code{cname}. Conventionally, the real and
imaginary components are prefixed with the letters \code{R} and \code{I}
respectively, while the position ($i,j$) in the matrix is referred to by a suffix
\code{ixj}. A complex parameter comes with a prefix~\code{C}.

Once an input file is read, an instance of the \code{SLHA.Card} class that can
be accessed via the \code{card} member of the basis class is created
and populated with the information provided as input. The content of the
mandatory \code{MASS} and \code{SMINPUTS} blocks is exported to data members of
the basis class named \code{mass} and \code{inputs} that can then be used
for accessing the SM parameters, while the CKM matrix is stored into the
\code{ckm} container of the basis class. In the \rosetta\ framework, the EFT
operator
coefficients are implemented as elements of the relevant basis class and can be
accessed via standard {\sc Python} methods. For instance, all the coefficients
associated with a basis object named \code{MyBasis} could be printed, together
with their values, by coding
\begin{verbatim}
  for k, v in MyBasis.items():
    print k, v
\end{verbatim}
In addition, a direct accessor to each EFT operator coefficient is created
from its name, which facilitates the implementation of the translation functions
that in general extensively reference individual parameter values. This however
assumes that there are no duplicate parameter names in the SLHA-like input file,
which nevertheless leads to a program exception. There are hence multiple ways
to access a given parameter. For example, a parameter \code{A} stored as the
third element of a block \code{MyBlock} that is part of the definition of a
basis \code{MyBasis} could be equally accessed as
\begin{verbatim}
  MyBasis['A']
  MyBasis.card['A']
  MyBasis.card.blocks['MyBlock']['A']
  MyBasis.card.blocks['MyBlock'][3]
\end{verbatim}
In the lines above, the parameter \code{A} is respectively accessed from the
\code{MyBasis} object, from the \code{SLHA.Card} instance associated with the
current basis and from the \code{SLHA.NamedBlock} object associated with the
\code{MyBlock} block (using either the parameter name or the counter as an
index).
\subsection{Implementing a new basis \label{s:exbasis}}
One of the important features of \rosetta\ is the intended ease with which a
user can define a new basis class to suit his/her specific physics needs. In the
context of an ultraviolet complete model, he/she may be interested in the
phenomenological consequences of a particular high-scale scenario in the EFT
framework. Imagining that he/she has derived all dimension-six Wilson
coefficients in a particular basis, \rosetta\ could be used to map these
coefficients to the \feynrules\ effective Lagrangian implementation in the
mass-eigenstate basis so that the collider
phenomenology of such a scenario could be investigated. This task is realized by
implementing a new basis in \rosetta\ and by connecting the new basis input parameters
to those of one of the existing core basis implementations.

Alternatively, the user may have developed a particular resource performing a
useful analysis or calculation in a non-standard basis choice. The corresponding
basis implementation in \rosetta\ with a translation to one of the core bases
could then allow one to use this tool in the context of all other existing basis
implementations in \rosetta\ and therefore greatly widen its scope. The
e{\sc HDecay} feature of \rosetta\ is an example of this, as it works with a set
of operators corresponding to the SILH basis. Via \rosetta, e{\sc HDecay} is now
available for calculations in the SILH, Warsaw and Higgs bases, as well as in
any additional basis that may be implemented in the future.

In this section, we provide an example that outlines the basic requirements for
implementing a new basis in \rosetta. We also refer the reader to the file
\code{Rosetta/TemplateBasis.py} which serves as a concrete toy example that can
be used as a template for creating a new basis class as well as the core basis implementations for more complete realizations.

All \rosetta\ basis classes inherit from the mother class \code{Basis}
implemented in the \code{Rosetta/internal/Basis.py} file.
This class contains all the machinery
necessary for reading, writing and translating so that a new basis
implementation solely demands the user to create a new file that must be saved
in the \code{Rosetta} directory and that includes the declaration of a
\code{Basis} subclass. The user has then to define the class attributes
described in Section~\ref{s:struct}. First, it is essential that the name of the basis class matches the filename in which it is saved minus the extension in order to
ensure a proper running of the program. Second, the \code{independent},
\code{blocks} and \code{flavored}
attributes of the class define the input parameters of the basis
and their desired SLHA-like structure, while the \code{required\_inputs} and
\code{required\_masses} lists are specified according to the needs of
the translation functions that are planned to be implemented. One can also specify a \code{dependent} attribute to explicitly define a particular parameter as dependent. For instance, the
following code refers to the implementation of a new basis class called
\code{MyBasis} and has been included in the file \code{Rosetta/MyBasis.py}.
\begin{verbatim}
  from internal import Basis
  class MyBasis(Basis.Basis):
    name = 'mybasis'
    independent=['A','B','one','two','MYxMAT']
    dependent = ['Cmat3x3']
    blocks = {'letters':['A','B','C'],
        'numbers':['one','two','three']}
    flavored = {'MYxMAT':{'kind':'hermitian',
                           'domain':'complex',
                           'cname':'mat'}}
    required_inputs = {1,2,4}
    required_masses = {24,25,6}
\end{verbatim}
This snippet of code specifies the declaration of the basis class \code{MyBasis}
whose unique string identifier is given by \code{mybasis}. The independent
parameters to be read from an input SLHA-like file are defined to be \code{A},
\code{B}, \code{one} and \code{two} and are assumed to be organized into the
two SLHA blocks \code{LETTERS} and \code{NUMBERS}. A flavored matrix,
\code{MYxMAT}, is also present and deemed to be an independent input parameter
\emph{except} for its (3,3) component that is explicitly included within the
\code{de\-pen\-dent} attribute of the basis class. The translation methods to be
implemented require the knowledge of six SM masses and parameters that must be
specified via the \code{required\_inputs} and \code{required\_masses} attributes
of the basis class. In our case, the electroweak inputs $\alpha^{-1}$, $G_F$ and
$m_Z$ are connected to the \code{SMINPUTS} block of the SLHA-like input
structure, while the $W$-boson, Higgs boson and top quark masses are connected
to its \code{MASS} block. The extra parameters \code{C} and \code{three} are
dependent parameters that the user has to define in terms of the independent and
SM parameters (see below). The non-SM part of an illustrative input file could
be
\begin{verbatim}
  BLOCK BASIS
   1 mybasis # input basis
  BLOCK LETTERS
   1 8.6e-2 # A
   2 0.002  # B
  BLOCK NUMBERS
   1 1.5e-2 # one
   2 2.8e-3 # two
  BLOCK MYxMAT
   1 1 3.4e-2 # Rmat1x1
   1 2 7.8e-5 # Rmat1x2
   1 3 5.2e-4 # Rmat1x3
   2 2 5.6e-3 # Rmat2x2
   2 3 3.3e-3 # Rmat2x3
  BLOCK IMMYxMAT
   1 2 9.9e-3 # Imat1x2
   1 3 1.9e-4 # Imat1x3
   2 3 4.6e-3 # Imat2x3
\end{verbatim}
while its SM part would include the \code{SMINPUTS} and \code{MASS} blocks with
values for the six above-mentioned SM inputs, as well as the two blocks related
to the CKM matrix in the case where one would be interested in using non-default
values for its elements. Only the relevant elements of \code{MYxMAT} need be
provided given that it is declared to be Hermitian, and the (3,3) element is left
unspecified as it is a dependent parameter.

The dependent parameters are evaluated via a method named
\code{calculate\_dependent()} that must be provided by the user. Continuing with
the example above, we include in the new basis class implementation the code
\begin{verbatim}
  def calculate_dependent(self):
    self['C']=(self['A']+self['B'])/2.
    self['three']=(self['one']-self['two'])/2.
    self['MYxMAT'][3,3]=10.*self['MYxMAT'][2,2]
\end{verbatim}
This imposes that the \code{C} parameter is defined as the mean of the \code{A}
and \code{B} parameters, that the \code{three} parameter equals half of the
difference of the \code{one} and \code{two} parameters and that the (3,3) entry
of the \code{MYxMAT} matrix is equal to 10 times the value of its (2,2) entry.

When executed, \rosetta\ begins with the reading of the input file and next
calls the \code{calculate\_dependent()} method for evaluating the remaining
basis parameters. \rosetta\ finally performs the translation to another basis by
using the translation methods defined by the user. Their implementation requires
the use of a \code{translation} decorator with an argument that refers to the
name of the target basis and that must match a basis implementation contained in
the \code{Rosetta} directory. For example, we could link the \code{mybasis}
basis above to the Warsaw basis by implementing
\begin{verbatim}
  @Basis.translation('warsaw')
  def mytranslation(self, wbasis):
    a_EW = 1./self.inputs['aEWM1']
    wbasis['cWW'] = a_EW*self['C']
    wbasis['WBxHpl'][1,1] = self['two']
    return wbasis
\end{verbatim}
Translation functions such as the \code{mytranslation(...)} one above take an
instance of the target basis class as their only argument and return it after
setting its parameter values. Relations involving (matrix) parameters with a
flavor structure should be performed in a flavor general way, as discussed in
Section~\ref{ss:flavour}.

If modifications to the SM input parameters need to be made (\textit{i.e.}, the
\code{mass} and \code{inputs} attributes of the basis class), the function
\code{modify\_inputs()} must be implemented similarly to the
\code{calculate\_dependent()} method. The following example defines a shift of
the $W$-boson mass by the \code{A} parameter,
\begin{verbatim}
  def modify_inputs(self):
    self.mass[24] = self.mass[24] + self['A']
\end{verbatim}
In general, the user-defined functions may require the evaluation
of parameters such as the weak and hypercharge gauge couplings or the
electroweak mixing angle. The choice of relations (\emph{e.g.}, tree-
or loop-level) to be used to consistently derive these parameters from the
inputs is left to the user. In the core bases provided
with \rosetta, the \code{calculate\_inputs() }$\,$ method relies on tree-level
relations to deduce all the SM parameters.

Having defined a basis class according to these specifications, \rosetta\ is
able to detect the presence of the basis implementation and to automatically
construct possible translation paths to other existing bases from the
user-defined translation functions. The recognition of the implemented basis by
\rosetta\ is also reflected in the help message accompanying the
\code{translate} script, the name of the new basis appearing as one of the
possibilities for the target basis option.

\subsection{Additional features}
\subsubsection{Flavor structure of the fermionic operators\label{ss:flavour}}
In the general case, each matrix block of the input file includes one entry for
each possible flavor assignment of the corresponding operator. The flavor
option of the \code{translate} executable introduced in
Section~\ref{s:commandline} allows the user to make assumptions on the flavor
structure of the operators so that \rosetta\ reads input files and generates
output files with a simplified block structure (unless the BSMC basis
is used as it requires all coefficients to be specified). Setting this option will act on \emph{all} of the matrix parameters declared in the \code{flavored} attribute of a basis class implementation. The flavor option can be
fixed either to \code{universal} where all matrices of operator coefficients are
proportional to the identity or to \code{diagonal} where only their
flavor-diagonal elements are retained.
In the \code{universal} case one is allowed to
define matrix blocks containing only the (1,1) element while in the
diagonal case, all three diagonal elements must be provided. Sample input files
can be found in the \code{Cards} directory of the program. In the definitions of the three core bases, the flavor-symmetry-breaking Yukawa-like operators are 
normalized by the masses of the fermions such that the \code{universal} flavor
option will lead to a minimally flavor-violating (MFV)
structure where the physical effects of the coefficients are scaled by the
corresponding fermion masses~\cite{D'Ambrosio:2002ex}. For example, in the Higgs basis, these Yukawa-like terms are written as:
\begin{align}
    \Delta\mathcal{L}_{\rm Yuk}=\sqrt{m_f^i m_f^j}\delta y_f^{ij}\bar{f}^i\left(\cos\phi^{ij}-i\gamma_5\sin\phi^{ij}\right)f^j.
\end{align}
The corresponding normalizations are also used for the Warsaw and SILH basis implementations in \rosetta\ to simplify the translations and also the possibility of encoding MFV into any EFT description. The same argument applies to the dipole operators, $O_{fV}$, as well as the $O_{Hud}$ operator mentioned in Section~\ref{s:input}. The former set of operators breaks the flavor symmetry in an identical way to the Yukawa-like operators and will hence receive the same $\sqrt{m_f^i m_f^j}$ normalization. In the latter case, the flavor structure of the operator requires two Yukawa insertions as it is composed of right-handed quarks only.
Moreover, being a charged-current operator, the MFV construction requires the insertion of the CKM matrix such that the operator is normalized as
\be
  {\cal O}_{Hud} = - i m_u^i m_d^j V^{ij}_{CKM}\big[\bar u^i\gamma^\mu d^j\big]
      \big[\tilde\Phi^\dag D_\mu\Phi\big]\ .
\ee
Since this particular operator is unique and maps to a single operator in all of the other core bases, the corresponding translations remain unaffected.
This normalization is however not the same as the one chosen in Ref.~\cite{HB}.
Users should therefore bear in mind these normalizations which have been chosen to single out operators that explicity break the flavor symmetry of the Lagrangian. That being said, they are merely a convenient way for the user to implement MFV and can be worked around if the user so desires.

\rosetta\ recognizes coefficients by their names so that the naming of the
elements of the matrix coefficients must respect the conventions described in
the previous section for their real (an \code{R} prefix) and imaginary (an
\code{I} prefix) parts, and for their position ($i,j$) inside the matrix
(an \code{ixj} suffix). Implementing
translations from flavored parameters should ideally always be done in the most
general case such that the various flavor options work correctly. To this aim,
basic matrix algebra operations have been implemeted in the
\code{internal/matrices.py} module of \rosetta. The available functions are
\code{matrix\_mult}, \code{matrix\_add}, \code{matrix\_sub} and
\code{matrix\_eq} and correspond to matrix multiplication, addition, subtraction
and assignment respectively. They can be used to assign values to a matrix SLHA
block according to the result of a specific operation between two other matrix
SLHA blocks. These functions require two mandatory arguments for the objects
between which the operation should be performed and a third optional argument
specifying the matrix block to which the result of the operation should be
assigned. For instance, \mbox{\code{matrix\_mult(M1,M2,M3)}} assigns to the matrix
\code{M3} the result of the multiplication of the matrices \code{M1} and
\code{M2}. If the \code{M3} argument is omitted, a generic matrix object is
returned such that matrix utility functions can be combined. The
\code{matrix\_eq(M1,M2)} method is the only exception. It takes two mandatory
arguments \code{M1} and \code{M2} and allows for the assignment of the elements
of the \code{M1} matrix to the \code{M2} matrix.

A concrete example can be found in the \code{calculate\_de\-pendent()} function
included in the Higgs basis implementation. The deviations of the $W$-boson
couplings to the weak doublet of left-handed quark fields $\delta g_L^{W_q}$ are
related to those of the $Z$-boson couplings to the individual left-handed
up-type and down-type quarks $\delta g_L^{Z_u}$ and $\delta g_L^{Z_d}$ via the
CKM matrix~$V_{\rm CKM}$,
\begin{align}
  \delta g_L^{W_q} = \delta g_L^{Z_u}\cdot V_{\rm CKM} -
     V_{\rm CKM}\cdot\delta g_L^{Z_d}\ .
\end{align}
The \rosetta\ implementation of this relation makes use of a combination of the
\code{matrix\_mult} and \code{matrix\_sub} method,
\begin{verbatim}
  matrix_sub(
    matrix_mult(HB['HBxdGLzu'], HB.ckm),
    matrix_mult(HB.ckm, HB['HBxdGLzd']),
    HB['HBxdGLwq']
  )
\end{verbatim}
where \code{HB} is an instance of the Higgs basis class. The third argument of
the \code{matrix\_sub} method allows one to assign the result of the matrix
subtraction to the elements of the \code{HBxdGLwq} matrix block. Matrix blocks also come with the \code{T()} and \code{dag()} methods for transposing and Hermitian conjugation.

\subsubsection{Interface to e{\sc HDecay}\label{ss:ehdecay}}
In order to calculate dimension-six operator contributions to the Higgs boson
width and branching ratios, \rosetta\ includes an interface to the e{\sc HDecay}
program~\cite{Contino:2013kra}. It can be switched on by executing the
\code{translate} script with the e{\sc HDecay} option (see
Section~\ref{s:commandline}). In order to use this feature, the path to a local
installation of e{\sc HDecay} on the user system should be specified in the
\code{Rosetta/config.txt} file, next to the \code{eHDECAY\_dir} keyword, and a
(possibly indirect) translation linking the basis of interest to the SILH basis
should exist. If so, the translation will be performed, e{\sc HDecay} will be
run internally and an SLHA decay block for the Higgs boson will be appended to
the output file.

Since the SILH basis description adopted in e{\sc HDecay} assumes the MFV
paradigm, an additional layer of translation is internally performed by
\rosetta\ to render its internal SILH basis implementation MFV-compliant.
Details can be found in Section~\ref{ss:yukawa}.

\section{Mapping different EFT basis choices}\label{sec:examples}
In this section, we discuss the BSMC Lagrangian containing redundant
parameters that is the default basis which \rosetta\ has been designed to
translate into. We explain the relations with the non-redundant Higgs, Warsaw
and SILH bases and focus on a subset of operators connected to single Higgs production at the LHC to provide examples of usage of \rosetta.

\subsection{The BSMC Lagrangian and the Higgs basis}\label{s:basis}

To study the Higgs boson properties in detail at the next LHC runs, the
LHCHXSWG has
proposed a parameterization of anomalous interactions of the SM fermions,
gauge bosons and the Higgs boson allowing both for a transparent linking
to physical observables and for an easy implementation in Monte-Carlo
event generators~\cite{HB}.  
The framework is that of a general effective Lagrangian defined in the mass-eigenstate basis,
where all kinetic terms are canonically normalized and diagonal, and where all mass terms are diagonal.
Moreover, the tree-level relations between the gauge couplings and the usual
electroweak input parameters ($G_F$, $\alpha(0)$, $m_Z$) are the same as in the SM.  
In such a frame, \textit{i.e.}, in the BSMC Lagrangian, the coefficients of the interaction terms in the Lagrangian are related in an intuitive way to quantities observable in experiment,
and any parameter in the effective Lagrangian can be measured.

The BSMC Lagrangian captures all physics effects that may arise in the presence of
lepton-number and baryon-number conserving dimension-six operators beyond the SM. 
However, it is more general than a basis defined before electroweak
symmetry breaking as it contains more free parameters.
This is because the $SU(3)_C\times SU(2)_L\times U(1)_Y$ gauge
symmetry {\em linearly} realized at the level of an operator basis
implies relations between different couplings of the effective
Lagrangian defined after electroweak symmetry breaking. The latter indeed only
respects the $SU(3)_C\times U(1)_{EM}$ symmetry.
In particular, the charged and neutral gauge boson interactions are related, as are those
with zero, one and two Higgs bosons.
These relations are {\em not} implemented at the level of the BSMC Lagrangian
so that
it may be used to study more general theories such as when the
electroweak symmetry
is non-linearly realized or when some operators of dimension greater than six
are included.

The {\em Higgs} basis has been proposed as a convenient
parameterization of another non-redundant dimension-six EFT basis. In this approach, the relations (that
hold in any non-redundant dimension-six basis of EFT operators) between different couplings of
the BSMC Lagrangian required by a linearly realized $SU(2)_L \times U(1)_Y$ local
symmetry are enforced. 
Furthermore, the Higgs basis has been defined by choosing a specific subset
of independent parameters from all couplings of the BSMC Lagrangian. 
The choice of the independent couplings is motivated by their direct
connection to observables constrained by electroweak precision and Higgs studies.
This approach is similar to the one introduced in Ref.~\cite{Gupta:2014rxa},
except that a different subset of couplings has been chosen, and
the number of independent couplings is the same as for any basis of
non-redundant dimension-six operators.
Moreover, there exists a linear one-to-one invertible transformation between
the independent couplings of the Higgs basis and the Wilson coefficients in any basis.
The remaining BSMC Lagrangian couplings are all dependent parameters
that can be expressed
in terms of the independent ones.

The BSMC Lagrangian is displayed in Ref.~\cite{HB}, up to four-fermion terms and
interactions involving five or more fields.
Here, to illustrate the relationship between the BSMC
and other bases,
we focus on a part of the Lagrangian describing the $CP$-even interactions of the Higgs boson with
two SM gauge bosons. After denoting by $G_\mu^a$, $W^\pm_\mu$, $Z_\mu$,
$A_\mu$ and $h$ the gluon, the $W$-boson, the $Z$-boson, the photon and the Higgs boson fields, respectively,
the relevant part of the Lagrangian reads
\begin{align}
 &\Delta{\cal L}_{h} = \frac{h}{v}\Big[
   2 \delta c_w m_W^2\, W^+_\mu W^{-\mu}      
    +\delta c_z m_Z^2\, Z_\mu Z^\mu \nn\\
   &\quad + c_{gg}\frac{g_s^2}{4}\, G_{\mu\nu}^aG^{\mu\nu}_a
    + c_{ww}\frac{g^2}{2}\, W^+_{\mu\nu}W^{-\mu\nu}
    + c_{zz}\frac{g^2}{4\ct^2}\, Z_{\mu\nu}Z^{\mu\nu} \nn\\
   &\quad + c_{z\gamma}\frac{gg'}{2}\, Z_{\mu\nu}A^{\mu\nu}
    + c_{\gamma\gamma}\frac{g'^2\ct^2}{4}\, A_{\mu\nu}A^{\mu\nu} \nn\\
   &\quad +c_{w\Box}g^2\, \big(W_\mu^-\partial_{\nu}W^{+\mu\nu} +{\rm h.c.}\big) 
    + c_{z\Box}g^2\,Z_{\mu}\partial_{\nu}Z^{\mu\nu} \nn\\
   &\quad + c_{\gamma\Box}gg'\, Z_{\mu}\partial_{\nu}A^{\mu\nu}
 \Big]\,.
\end{align}
In our notation, $\ct$ ($\st$) stands for the cosine (sine) of the electroweak
mixing angle, $v$ for the vacuum expectation value of the neutral component of
the Higgs doublet $\Phi$, and $g_s$, $g$ and $g'$ are the
strong, weak and hypercharge coupling constants. Moreover, we
have introduced the field strength tensors of the gauge bosons that we define as
\be\bsp
 V_{\mu\nu} &=\partial_{\mu}V_{\nu}-\partial_{\nu}V_{\mu}\quad\text{for}\quad
      V=W^{\pm}, Z \text{ and } A\,, \\
 G_{\mu\nu}^a &=\partial_{\mu}^{}G_{\nu}^a-\partial_{\nu}^{}G_{\mu}^a
    +g_s f^a{}_{bc} G_{\mu}^bG_{\nu}^c\,,
\esp\ee
in which $f^a{}_{bc}$ are the structure constants of $SU(3)_C$. 

The Lagrangian above contains ten real coupling parameters that
are all independent in the BSMC picture.
However, if $\Delta{\cal L}_{h}$ originates from an EFT with dimension-six operators, only six
of these couplings are independent and the remaining four can be expressed in terms
of these six and of the SM parameters.
In the Higgs basis,
$\delta c_z$, $c_{gg}$, $c_{zz}$, $c_{z\gamma}$, $c_{\gamma\gamma}$ and $c_{z\Box}$
are chosen as independent parameters
and the four remaining couplings are calculated as
\be\bsp
 \delta c_w &= \delta c_z +4\delta m\,, \\
 c_{ww} &= c_{zz} +2\st^2 c_{z\gamma} +\st^4 c_{\gamma\gamma}\,, \\
 c_{w\Box} &= \frac{g^2c_{z\Box}+g'^2c_{zz}
     -(g^2-g'^2)\st^2c_{z\gamma}-g^2\st^4c_{\gamma\gamma}}{g^2-g'^2}\,, \\
 c_{\gamma\Box} &= \frac{2g^2c_{z\Box}+(g^2+g'^2)c_{zz}
    -(g^2-g'^2)c_{z\gamma}-g^2\st^2c_{\gamma\gamma}}{g^2-g'^2}\,.
  \label{relationsh}
\esp\ee
In the first of these relations, $\delta m$ denotes the shift of the $W$-boson
mass that is possibly induced by the presence of higher-dimensional operators
and that we normalize as
\be
\Delta{\cal L}_{\rm mass} = 2\delta m\frac{g^2 v^2}{4}\, W_\mu^+ W^{-\mu} \,.
\ee

An input file (that we name \code{HiggsBasis.dat} in our example) describing
this part of the Higgs basis Lagrangian would be of the form
\begin{verbatim}
  BLOCK BASIS
    1  higgs # basis
  BLOCK HBxh
    1  1.00000e-01 # dCz
    2  1.00000e-01 # Cgg
    3  1.00000e-01 # Caa
    4  1.00000e-01 # Cza
    5  1.00000e-01 # Czz
    6  1.00000e-01 # Czbx
\end{verbatim}
It includes, in addition to the blocks above, the SM parameters as well as
vanishing values for all other EFT coefficients. 
In order to export this setup
to the BSMC Lagrangian, we use \rosetta\ by typing in a shell
\begin{verbatim}
  ./translate HiggsBasis.dat
\end{verbatim}
\rosetta\ first calculates all dependent coefficients and next generates an
output file named \code{HiggsBasis\_new.dat} given in the framework of
the BSMC Lagrangian.
This file contains in particular values for the four $\delta c_w$, $c_{ww}$, $c_{w\Box}$
and $c_{\gamma\Box}$ dependent parameters, the corresponding output block being, according to Eq.~\eqref{relationsh},
\begin{verbatim}
  BLOCK BCxh 
    1  +1.00000e-01 # dCw
    1  +1.00000e-01 # dCw
    2  +1.00000e-01 # dCz
    3  +1.52190e-01 # Cww
    4  +1.00000e-01 # Cgg
    5  +1.00000e-01 # Caa
    6  +1.00000e-01 # Cza
    7  +1.00000e-01 # Czz
    8  +1.56506e-01 # Cwbx
    9  +1.00000e-01 # Czbx
    10 +3.41838e-01 # Cabx
\end{verbatim}
In addition, the \code{HiggsBasis\_new.dat} file also includes extra
non-vanishing coefficients that are linked to the six independent parameters
$\delta c_z$, $c_{gg}$, $c_{\gamma\gamma}$, $c_{z\gamma}$, $c_{zz}$ and
$c_{z\Box}$ by gauge invariance. For instance, a di-Higgs coupling to two
gluonic field strength tensors is present,
\begin{verbatim}
  BLOCK BCxhh
    4   1.00000e-01 # cgg2
\end{verbatim}

\subsection{The Warsaw basis}

The ten interaction terms of the $\Delta{\cal L}_{h}$ Lagrangian introduced in
Section~\ref{s:basis} can be seen as generated by six independent operators of
the Warsaw basis,
\begin{align}
 &\Delta{\cal L}_{h}^{\rm W} =\frac{1}{v^2}\Big[
      c_{GG}\frac{g_s^2}{4}\, \Phi^\dag \Phi\, G_{\mu\nu}^aG^{\mu\nu}_a
    + c_{WW}\frac{g^2}{4}\, \Phi^\dag \Phi\, W_{\mu\nu}^iW^{\mu\nu}_i \nn\\
  &\quad\, 
    + c_{WB}gg'\, \Phi^\dag \sigma_i \Phi\, W_{\mu\nu}^iB^{\mu\nu} 
    + c_{BB}\frac{g'^2}{4}\, \Phi^\dag \Phi\, B_{\mu\nu}B^{\mu\nu} \nn\\
  &\quad\,  
    + c_H\, \partial_\mu\big[\Phi^\dag \Phi\big]
         \partial^\mu\big[\Phi^\dag \Phi\big]
    + c_T \big[ \Phi^\dag {\overleftrightarrow{D}}_\mu \Phi \big]
          \big[ \Phi^\dag {\overleftrightarrow{D}}^\mu \Phi \big]\Big]\,.
\end{align}
In this expression, we have introduced the Pauli matrices $\sigma_i$, the
Hermitian derivative operator,
\be
  \Phi^\dag {\overleftrightarrow D}_\mu \Phi = 
    \Phi^\dag (D_\mu \Phi) - (D_\mu\Phi^\dag) \Phi \,,
\ee
the gauge-covariant derivative and the hypercharge and weak field
strength tensors
\be\bsp
 & D_\mu\Phi = \Big(\partial_\mu -\frac{i}{2} g \sigma_{k} W_\mu^k 
              - \frac{i}{2} g' B_\mu\Big) \Phi \,, \\
 & W_{\mu\nu}^i =\partial_{\mu}^{}W_{\nu}^i-\partial_{\nu}^{}W_{\mu}^i
    +g \epsilon^i{}_{jk} W_{\mu}^jW_{\nu}^k \,,\\
 & B_{\mu\nu} =\partial_{\mu}B_{\nu}-\partial_{\nu}B_{\mu} \,.
\esp\ee

The six Wilson coefficients $c_{GG}$, $c_{WW}$,
$c_{WB}$, $c_{BB}$, $c_H$ and $c_T$ appearing in $\Delta{\cal L}_{h}^{\rm W}$ are related to
the ten couplings in the effective Lagrangian $\Delta{\cal L}_{h}$ as
\begin{align}
  \nn\delta c_w =&\ -c_H - \frac{4 g^2g'^2}{g^2-g'^2} c_{WB}
    + \frac{4 g^2}{g^2-g'^2} c_T - \frac{3g^2+g'^2}{g^2-g'^2} \delta v\,,\\
  \nn\delta c_z =&\ -c_H - 3 \delta v \,,\\
  \nn c_{gg} =&\ c_{GG}\,,\\
  \nn c_{ww} =&\ c_{WW}\,,\\
  \nn c_{zz} =&\ \frac{g^4c_{WW}+4g^2g'^2c_{WB}+g'^4c_{BB}}{(g^2+g'^2)^2}\,,\\
  c_{z\gamma} =&\ \frac{g^2 c_{WW}-2(g^2-g'^2)c_{WB}-g'^2c_{BB}}{g^2+g'^2}\,,\\
  \nn c_{\gamma\gamma} =&\ c_{WW} + c_{BB} - 4 c_{WB}\,,\\
  \nn c_{w\Box} =&\ \frac{2}{g^2-g'^2} \big[ g'^2 c_{WB} - c_T + \delta v\big]\,,\\ 
  \nn c_{z\Box} =&\ -\frac{2}{g^2} \big[ c_T - \delta v\big]\,,\\
  \nn c_{\gamma\Box} =&\ \frac{2}{g^2-g'^2} \big[ (g^2+g'^2) c_{WB} -2 c_T
     + 2 \delta v\big]\,.
\end{align}
Here, $\delta v$ is defined by
\be
  \delta v = \frac12 \Big[(c'_{H\ell})_{11} + (c'_{H\ell})_{22}\Big]
     - \frac14 (c_{\ell\ell})_{1221}
\ee
and summarizes the dependence on the additional Warsaw basis operators,\footnote{%
These operators contribute to the muon decay at the tree level.
Taking this into account  leads to a shift between the measured Fermi constant  and  the vacuum expectation value of the Higgs field, which motivates the notation $\delta v$.}
\be
    \frac{i}{v^2} c'_{H\ell}\, \big[\bar \ell\sigma_i\gamma_\mu \ell\big]
       \big[\Phi^\dag \sigma^i {\overleftrightarrow D}^\mu \Phi \big]
  + \frac{1}{v^2} c_{\ell\ell}\, \big[\bar\ell\gamma_\mu \ell\big]
         \big[\bar\ell\gamma^\mu \ell\big]\,.
\ee

Starting from the Higgs basis example of Section~\ref{s:basis} where all
independent parameters are fixed to 0.1, we employ \rosetta\ to invert the
relations of this section and calculate the numerical values of the Warsaw basis
coefficients included in $\Delta{\cal L}_{h}^{\rm W}$ that would yield the same
$\Delta{\cal L}_{h}$ Lagrangian. Typing in a shell
\begin{verbatim}
 ./translate HiggsBasis.dat -t warsaw
\end{verbatim}
we obtain an output file where several non-zero EFT coefficients can be found.
The numerical value of those on which we focus here can be extracted from the
generated file,
\begin{verbatim}
  BLOCK BASIS
    1   warsaw # translated basis
  BLOCK WBxH4D2 # 
    1    -1.98704e-01 # cH
    2    +1.18790e-02 # cT
  BLOCK WBxV2H2 # 
    1    +1.00000e-01 # cGG
    2    +1.52190e-01 # cWW
    3    +5.45943e-03 # cBB
    4    +1.44124e-02 # cWB
\end{verbatim}
In our benchmark scenario, the $\delta v$ shift is vanishing.

\subsection{SILH basis}

We now consider the case where all operators included in the $\Delta{\cal L}_{h}$
Lagrangian of Section~\ref{s:basis} are induced by a set of operators of the
SILH basis,
\begin{align}
  &\Delta{\cal L}_{h}^{\rm S} = \frac{1}{v^2} \Big[
     s_{GG}\frac{g_s^2}{4}\, \Phi^\dag \Phi\, G_{\mu\nu}^aG^{\mu\nu}_a
   + s_{BB}\frac{g'^2}{4}\,  \Phi^\dag \Phi\, B_{\mu\nu}B^{\mu\nu} \nn\\
 &\quad
   + s_W\frac{ig}{2}
       \big[\Phi^\dag\sigma^i{\overleftrightarrow D}_\mu\Phi\big]
       \partial_\nu W^{\mu\nu}_i
   + s_B\frac{ig'}{2} \big[\Phi^\dag {\overleftrightarrow D}_\mu \Phi \big]
       \partial_\nu B^{\mu\nu} \nn\\
 &\quad
   + i\,s_{HW}g\, \big[D_\mu\Phi^\dag \sigma^i D_\nu\Phi\big]
 W^{\mu\nu}_i 
   + i\,s_{HB}g'\, \big[D_\mu\Phi^\dag D_\nu\Phi\big] B^{\mu\nu} \nn\\
 &\quad
   + s_{2W}\, D^\mu W_{\mu\nu}^i D_\rho W^{\rho\nu}_i
   + s_{2B}\, \partial^\mu B_{\mu\nu} \partial_\rho B^{\rho\nu} \nn\\
  &\quad
   + s_H\, \partial_\mu\big[\Phi^\dag \Phi\big]
       \partial^\mu\big[\Phi^\dag \Phi\big]
   + s_T\, \big[ \Phi^\dag {\overleftrightarrow{D}}_\mu \Phi \big]
         \big[ \Phi^\dag {\overleftrightarrow{D}}^\mu \Phi \big] \nn\\
 &\quad
   + i\, s'_{H\ell}\, \big[\bar \ell\sigma_i\gamma_\mu \ell\big]
       \big[\Phi^\dag \sigma^i {\overleftrightarrow D}^\mu \Phi \big]
  \Big] \,.
\end{align}

The ten Wilson coefficients included in $\Delta{\cal L}_{h}$ can be rewritten in
terms of the eleven parameters appearing in $\Delta{\cal L}_{h}^{\rm S}$ as
\be\nn\bsp
  \delta c_w =&\ -s_H
      -\!\frac{g^2g'^2 (s_W\!+\!s_B\!+\!s_{2W}\!+\!s_{2B})}{g^2-g'^2}
      \!-\! \frac{4g^2}{g^2-g'^2} s_T\phantom{aaaaaaa}\\&\qquad
      \!+\! \frac{3g^2+g'^2}{g^2-g'^2}\delta v\,,\\
  \delta c_z =&\ -s_H - 3 \delta v \,,\\
  c_{gg} =&\ s_{GG}\,,\\
  c_{ww} =&\ -s_{WW}\,,\\
  c_{zz} =&\ -\frac{g^2 s_{HW}+g'^2s_{HB}-g'^2\st^2
  s_{BB}}{g^2+g'^2}\,,\\
\esp\ee
\be\bsp \hspace{-.4cm}c_{z\gamma} = -\frac{s_{HW}-s_{HB}}{2} - \st^2 s_{BB}\,,\\ 
\esp\ee\be\bsp\nn
  c_{\gamma\gamma} =&\ s_{BB}\,,\\
  c_{w\Box} =&\ \frac12 s_{HW} + \frac{g^2 (s_W+s_{2W}) + g'^2 (s_B+s_{2B})
      - 4 s_T + 4 \delta v}{2(g^2-g'^2)} \,,\\
  c_{z\Box} =&\ \frac{g^2 (s_W\!+\!s_{2W}\!+\!s_{HW})
      \!+\! g'^2 (s_B\!+\!s_{2B}\!+\!s_{HB}) \!-\! 4 s_T
      \!+\! 4 \delta v}{2g^2}\,,\\
  c_{\gamma\Box} =&\ \frac{s_{HW}\!-\!s_{HB}}{2}\!+\!\frac{g^2 (s_W\!+\!s_{2W})
     \!+\! g'^2 (s_B\!+\!s_{2B}) \!-\! 4 s_T \!+\! 4 \delta v}{g^2-g'^2}\,,
\esp\ee
where $\delta v =\frac12 (s'_{H\ell})_{22}$. \rosetta\ can be used to extract
the numerical values of the independent SILH parameters by inverting the above
relations. Adopting the benchmark scenario of Section~\ref{s:basis} where all
the relevant Higgs basis independent parameters have been fixed to 0.1, we type
in a shell
\begin{verbatim}
  ./translate HiggsBasis.dat -t silh
\end{verbatim}
so that we can extract all the required SILH coefficients from the generated
output file,
\begin{verbatim}
  BLOCK BASIS
    1   silh # translated basis
  BLOCK SBxH4D2 # 
    1    -1.00000e-01 # sH
    2    +0.00000e+00 # sT
  BLOCK SBxV2H2 # 
    1    +1.00000e-01 # sGG
    2    +1.00000e-01 # sBB
    3    +4.65203e-01 # sW
    4    -4.65203e-01 # sB
    5    -1.52190e-01 # sHW
    6    +9.45406e-02 # sHB
    7    +0.00000e+00 # s2W
    8    +0.00000e+00 # s2B
\end{verbatim}

\subsection{Yukawa-like operators\label{ss:yukawa}}
An important difference between the definitions of the SILH and Warsaw bases
provided in the LHCHXSWG proposal~\cite{HB}
and their original descriptions lies in the forms of the Yukawa-like operators,
\begin{align}
    \Delta\mathcal{L}_{\text{Yuk}}= \frac{(c_f)_{ij}}{v^2}\,\big(\Phi^\dagger\Phi\big)\,\bar{F}_L^{i}\Phi f_R^{j}\, ,
\end{align}
where $F_L$ and $f_R$ denote a generic weak doublet of left-handed fermions
and a generic weak singlet of right-handed fermions respectively. In the
original Warsaw basis definition, these Yu\-ka\-wa-like operators take the above form.
In the LHCHXSWG proposal (on which \rosetta\ is based), these operators have
been redefined in a way allowing one to decouple their contributions to the
fermion masses (that are extracted from appropriate measurements and thus fixed), as well as to simplify the implementation of MFV,
\begin{align}
    \Delta\mathcal{L}_{\text{Yuk}}^\prime = -\frac{\sqrt{m_i m_j}}{v}\frac{(c^\prime_f)_{ij}}{v^2}\,\Big(\Phi^\dagger\Phi - \frac{v^2}{2}\Big)\,\bar{F}_L^{\prime i}\Phi f_R^{\prime j}\, ,
\end{align}
where the primes denote fields taken in the mass eigenbasis. The Wilson
coefficients $c_f$ and $c'_f$ are related by unitary transformations $U_L$ and $U_R$ that map
the field gauge eigenbasis to the mass eigenbasis with the would-be mass
modifications absorbed into the diagonalized Yukawa matrices $Y^D_f$,
\be
    c^\prime_f = \frac{v}{\sqrt{m_i m_j}}U_L^\dagger c_f U_R \quad\text{and}\quad
    Y^D_f = U_L^\dagger Y_f U_R + \frac{c_f}{2}\, .
\ee
In the original SILH basis description, an additional assumption of minimal
flavor violation is included, such that the flavor structure is taken aligned
with the Yukawa matrices,
\begin{align}\label{e:SILHMFV}
    \Delta\mathcal{L}_{\text{Yuk}}^{\rm MFV}= (Y_f)_{ij}\,\frac{c^{\rm MFV}_f}{v^2}\,\big(\Phi^\dagger\Phi\big)\,\bar{F}_L^{i}\Phi f_R^{j}\, .
\end{align}
The Wilson coefficients $c_f^{\rm MFV}$ are therefore proportional to the
identity matrix in flavor space and are thus universal. Thanks to the convenient normalizations, they are now trivially related linearly to
those of the Warsaw and SILH basis descriptions of the LHCHXSWG proposal by
\be
   (c_f)_{ii}= \frac{v}{m_i}Y^D_f c_f^{\rm MFV}=\sqrt{2}c_f^{\rm MFV}\, .
\ee

These relations are used internally for the e{\sc HDecay} interface of \rosetta, which takes
SILH basis input parameters assuming the MFV
convention of Eq.~\eqref{e:SILHMFV}. In order to consistently use e{\sc HDecay},
\rosetta\ translates these coefficients from the alternative version of the SILH
basis detailed in Ref.~\cite{HB}. As a consequence, a general flavor structure
cannot be employed when making use of the e{\sc HDecay} interface. Although it
is in principle possible to input different values for the $c^{\rm MFV}_c$, $c^{\rm MFV}_b$,
$c^{\rm MFV}_t$, $c^{\rm MFV}_\mu$ and $c^{\rm MFV}_\tau$ parameters (referred in Ref.~\cite{Contino:2013kra} as $\bar{c}_c$, \textit{etc.}) when running e{\sc HDecay}
on its own, large deviations from non-universality in these coefficients
consist of a significant departure from the MFV paradigm and should not be used
for complete consistency within \rosetta.

\section{Summary} \label{sec:conclusions}
In this paper, we have introduced the \rosetta\ package, a {\sc Python} program
dedicated to the translation of a given EFT basis of independent operators to
other viable basis choices. We have also included, in
this document, technical details so that users can easily extend the program and
implement their own choices of EFT operator basis.

Currently, the program allows the user to translate
benchmarks designed in the Higgs, SILH and Warsaw bases into any of these three
bases. In addition, the code also expresses any scenario in terms of the BSMC
Lagrangian of EFT operators, a basis that has been defined from the Higgs basis after
ignoring all relations among the operators that are
induced by a linear realization of the
electroweak symmetry. A
\feynrules\ implementation allows  \rosetta\ to be linked to other
high-energy physics tools. The relations among the different Wilson coefficients
that hold in the context of the Higgs, SILH and Warsaw bases of independent
operators have been implemented into \rosetta\ so that they are preserved when
a setup is exported to the BSMC Lagrangian by the program.
This scheme has the strength to be easily generalizable to study different
setups providing a description of the Higgs boson properties, such as those
with a non-linearly
realized electroweak symmetry or including higher-dimensional operators
beyond dimension six.

In the
future, we believe that translations from one basis to another will allow for
broadening the scope and the use of past calculations very relevant for
precision Higgs physics. Along these lines, higher-order calculations in QCD
performed
in the BSMC Lagrangian~\cite{Maltoni:2013sma,Demartin:2014fia,Demartin:2015uha}
could be used within any given
EFT language, and the renormalization group running of the Wilson
coefficients, that has been calculated in the SILH basis \cite{Elias-Miro:2013gya,Elias-Miro:2013mua} and in the Warsaw
bases~\cite{Jenkins:2013zja,Jenkins:2013wua,Alonso:2013hga}, could be
exported to different bases too.

\section*{Acknowledgments}

The authors are grateful to Fabio Maltoni for lively interactions during all
phases of this project and would also like to thank Christophe Grojean, Alex Pomarol and Michael
Trott for useful comments. This work has been partially supported 
by the FP7 Marie Curie Initial Training Network MCnetITN
(PITN-GA-2012-315877),
by the Belgian Federal Science Policy Office through the Interuniversity
Attraction Pole P7/37, 
by the Strategic Research Program `High Energy Physics' and the Research
Council of the Vrije Universiteit Brussel,
by the Swiss National Science Foundation under the
Ambizione grant PZ00P2 136932,
and by the Theory-LHC-France initiative of the CNRS (INP and IN2P3).

\bibliography{library}
\bibliographystyle{JHEP}

\end{document}